# Beam Dynamics simulations for ERDC project – SRF linac for industrial use

N. Solyak, I. Gonin, , A. Saini, V. Yakovlev, APS-TD, Fermilab, Batavia, USA
C. Edwards, J.C.Thangaraj, IARC, Fermilab, R. Kostin, C. Jing, Euclid Techlabs LLC

*Abstract*

Compact conductively cooled SRF industrial linacs can provide unique parameters of the electron beam for industrial applications. (up to 10MeV, 1MW). For ERDC project we designed normal conducting RF injector with thermal RF gridded gun integrated in first cell of multi-cell cavities. For design of the RF gun we used MICHELLE software to simulate and optimize parameters of the beam. Output file was converted to ASTRA format and most beam dynamic simulations in multi-cell normal conducting cavity and cryomodule were performed by using ASTRA software. For cross-checking we compare results of MICHELLE and ASTRA in first few cells. At the end of injector beam reach ~250keV energy which allow to trap bunch in acceleration regime without losses in TESLA like 1.3 GHz cavity. Short solenoid at the end of injector allow to regulate transverse beam size in cryomodule to match beam to extraction system and also reduce charge losses in accelerator.

## INTRODUCTION

The new concept of a compact linear accelerator for industrial application suggested in [1] is based on use of Superconducting Radio-Frequency (SRF) cavities. Several proposals of the compact conduction cooled linac are get funded for R&D stage. Attractive electron beam parameters for industrial application is 10 MeV with average beam powers of 100's kW.

The Illinois Accelerator Research Center (IARC) of Fermilab has started design, construction, and validation of a compact 20 kW prototype of the linac capable of operating 10MeV, 2mA beam in a continuous wave (CW) mode (ERDC project). The linac employs a 9-cells 1.3GHz TESLA-type Nb$_3$Sn coated conduction-cooled cavity. As a source of electron beam we propose to use external injector capable to produce bunches extracted from gridded RF gun and accelerated in copper cavities up to 250-300 keV. The general design of the gun is reported earlier [2,3]. The engineering design of the cathode-grid unit has been done at HeatWave Labs, Inc. and two prototypes were fabricated

General layout of the linac, including RF injector with 6 cell cavity and cryostat with SC 9-cell cavity are presented in Figure 1. On the left side one can see resonator integrated with RF gun and providing RF voltage on the grid. Playing with RF amplitude and DC voltage on the grid and phase/amplitude of accelerating cell allows us to get required 2mA beam current and regulate bunch length, output energy and energy spread.
.

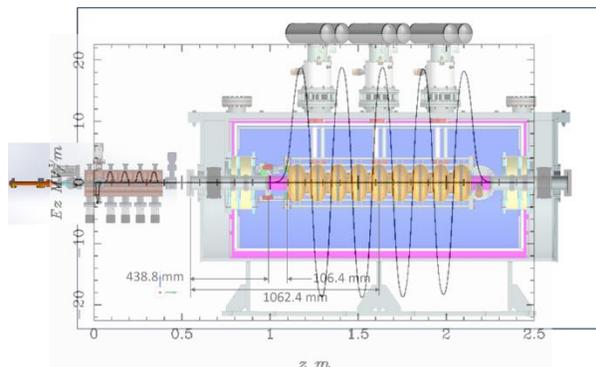

Figure 1: General layout of the 20kW accelerator with external injector and superconducting 9-cells 1.3 GHz cavity.

## RF GUN OPTIMIZATION

### MICHELLE simulation

The MICHELLE code [4] is a 3D electrostatic particle-in-cell (PIC) code that has been designed to address the beam optics modelling and simulation of charged-particle transport. Problem classes specifically targeted include depressed collectors, gridded-guns, multi-beam guns etc. Cross-checking MICHELLE with SMASON-2D code demonstrated excellent agreement. The MICHELLE approach uses space charge limited emission based on the Child-Langmuir law.

The beam dynamics simulation of the RF gun + injector cavity assembly was performed with 3D MICHELLE code. An RF voltage with a DC bias is applied to the cathode-grid gap to form the electron bunches. After emission, these bunches are captured and accelerated by RF field in the gun-cell (C0) of the injector. The phase shift φ between the RF fields in the gun and in the first cell is one of the optimization parameters. Other parameters are DC and RF voltages between cathode and grid.

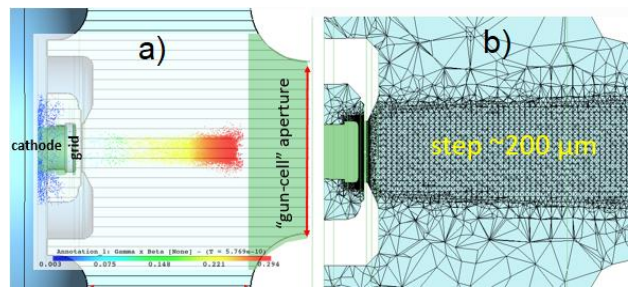

Figure 2: Example of MICHELLE simulation of RF gun. a) beam dynamics in the gun and accelerating cell. b) typical mesh-size: 25 μm-on the grid area and 200 μm in accelerating cell beam transportation area..

Figure 2 show geometry of the gun and cell and accelerated beam. For mesh size we chose fine mesh in cathode-grid area (25 μm) and near the axis of accelerating cell (200 μm). Grid parameters were optimize for current of 2 mA and output energy of ~30 keV.

MICHELLE simulations have been performed for three bunch length scenario, shown in Table 1 (V1-short bunch, V2-medium and V3- long bunch). This bunch length scenario was used to minimize beam losses during acceleration.

Table 1: Parameters of the accelerating cells for three bunch-length scenario.

| Parameter | V1 | V2 | V3 | Unit |
|---|---|---|---|---|
| Energy | 32 | 29.7 | 28.5 | keV |
| Voltage $_{DC}$ | 150 | 60 | 30 | V |
| Voltage $_{RF}$ | 200 | 100 | 65 | V |
| Cavity amplitude | 40 | 35 | 35 | kV |
| Cavity phase | 20 | 10 | 20 | deg |
| Beam $\phi$ duration | 2.1 | 4.7 | 5.7 | °rms |

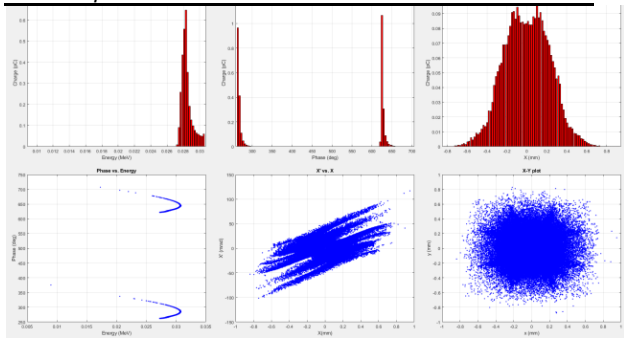

Figure 3: Beam distribution at the exit of gun cell for short bunch scenario (V1)

MICHELLE was also used to simulate beam dynamic in second RF cell (C1-16). This simulations is time consuming (taking into account mesh size, number of particles in bunch and number of bunches), that's why we used symmetry of gun and simulate ¼ of full geometry. We also used ASTRA to simulate beam dynamics in second cell using results from MICHELLE saved after first cell (C0) converted to the ASTRA format. Results of both codes are in a good agreement as one can see from Figure 4. This cross-checking convince us to use ASTRA code for whole system, except gridded cathode and first acceleration cell.

| MICHELLE/ASTRA parameters | | |
|---|---|---|
| $E_{av}$, keV | 75.3 | 75.32 |
| $\sigma_E$, keV | 1.74 | 1.69 |
| $\sigma_{phase}$, ° | 4.5 | 4.5 |

Figure 4: Comparison of beam parameters at the exit of second cell calculated by MICHELLE (left column) and ASTRA (right).

## INJECTOR

General layout of injector is shown in figure 3. Detailed information about injector optimization, mechanical design and thermal analysis is reported in this conference [5]. Voltage and RF power requirements for cells are shown in Table 1. Gridded gun installed in cell C0-8, second cell C1-16 and 4 regular cells C3-C6 accelerate the beam up to 250-300keV. Available RF per cell is limited by 800W CW.

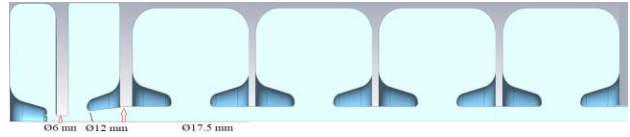

Figure 5: General layout of the injector

Table 2: Parameters of the accelerating cell, voltage and power

| Cell# Gap | C0-8 | C1-16 | C2-25 | C3-25 | C4-25 | C5-25 | total |
|---|---|---|---|---|---|---|---|
| V, KV | 35 | 40 | 50+ | 50+ | 50+ | 50+ | 275+ |
| P, W | 780 | 640 | 460 | 460 | 460 | 460 | 3300 |

## BEAM DYNAMIC IN ACCELERATOR

Energy gain in full injector (maximum ~250kV) and beam charge and power losses (12% charge, max power ~50W) calculated for short bunches (V1) are shown in Figure 6. Dots present geometry of the cells. Without focusing beam increases in size and start loose particles after 4 cell.

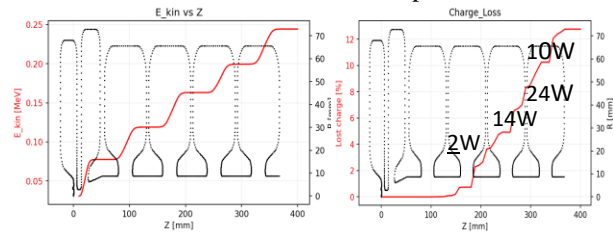

Figure 6: Beam Energy (left) and beam charge losses in injector for on-crest acceleration (right). RF amplitudes are shown in Table 2, phases: 20, 134, 75, -140, -6, -140 degrees.

ASTRA also simulated beam in whole accelerator, including 9-cell cavity in cryomodule as shown in Figure 7. Aperture (shown in blue) increase from 17.5 mm in injector (410mm and ~100mm space after) to 70mm in cryomodule. Full distance between last cell of injector and 1$^{st}$ cell of 9-cell cavity is ~645mm. In full system beam losses increased from 12% in injector only to ~26%, without losses in CM.

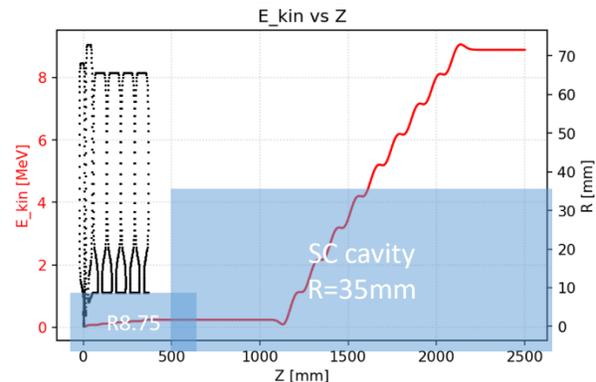

Figure 7: Energy gain in injector and 9-cell SC cavity,

We tried two ways to focus the beam and decrease beam losses:
- focusing solenoid right after injector
- RF focusing by shifting off-crest RF phases in injector cells.

For short bunch solenoid helps to reduce charge losses from 26 % to 16%. Table 3 summarizing charge losses for different bunch lengths and shows effect of solenoid focusing. One can see that bunch length is playing significant role in losses in system. For long bunch (V3) simulation predicts 3.7% losses in injector and 2% full losses with optimized solenoid field 17mT. In all cases there is no loss in CM observed.

Table 3: Charge losses (in %)

| Bunch length | Injector (%) | Full % | Sol. % |
| --- | --- | --- | --- |
| V1 | 12 | 26 | 16 |
| V2 | 4 | 6.5 | 4.5 |
| V3 | 3.7 | 13 | 2 |

Beam distribution for long bunch V3 along the accelerator are shown in Figure 8.

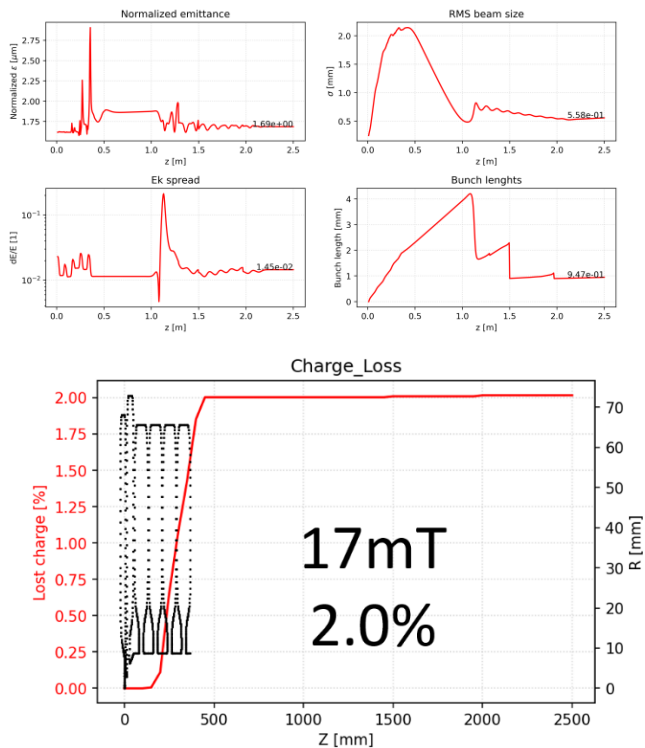

Figure 8: Normalized Emittance, RMS bunch size, Energy spread and bunch length are plotted along beamline (top). Charge losses with 17mT solenoid minimized charge losses.

## CONCLUSION

The Illinois Accelerator Research Center (IARC) of Fermilab has started design, construction, and validation of a compact 20 kW prototype of the linac capable of operating 10MeV, 2mA beam in a continuous wave (CW) mode (ERDC project). We propose to use external injection to generate bunched beam and accelerate it up to 250keV for injection in CM with 1.3GHz 9-sell cavity coated by Nb3Sn. MICHELLE 3D software was used for simulation of the beam dynamics in gridded RF gun and first accelerating cell. Beam propagation through the rest of injector and cryomodule were simulated by ASTRA code. Huge parameter space was studied to optimize injector design and significantly reduce charge losses in injector part. Solenoid between injector and cryomodule helps to reduce beam losses to 2% and optimize bunch distribution at the exit of cryomodule. Results of simulation is used for mechanical design of injector and gridded gun for the project.

## ACKNOWLEDGEMENTS

The experiments described and the resulting data presented were funded under PE 0603119A, Project B03 "Accelerator Technology for Ground Maneuver", managed by the US Army Engineer Research and Development Center. The work described in this presentation was conducted at Fermilab. Permission was granted by Fermilab and ERDC to publish this information."

## REFERENCES

[1] R.D. Kephart, et al, "SRF, Compact Accelerators for Industry and Society," 17th International Conference on RF Superconductivity, Whistler, September 13-18, 2015, FRBA03

[2] I.Gonin, et al., "Built-in thermionic electron source for an SRF linacs', IPAC21, THPAB156, pp4062-4064.

[3] I.Gonin et al, "Gridded RF gun design for an SRF linac applications", in this proceedings, MOPB062, LINAC24

[4] J.Petillo et al., "Application of the Finite-Element MICHELLE Beam Optics Code to RF Gun Modeling", in Proc. IVEC, Monterey, USA, Apr. 2006, pp. 83-84. doi:10.1109/IVELEC.2006.1666195

[5] R.Kostin et al, "CW copper injector for SRF industrial cryomodules", in this proceedings, THPB021, LINAC24